%% file: main.tex
\documentclass[reprint,
 amsmath,amssymb,
 aps,prd,
 superscriptaddress, floatfix,
 longbibliography
]{revtex4-2}

\usepackage{graphicx}
\usepackage{dcolumn}
\usepackage{lineno}
\usepackage{bm}
\usepackage{comment}
\usepackage{xspace}
\usepackage[dvipsnames]{xcolor}
\usepackage{units}
\usepackage{hyperref}
\usepackage{url}
\usepackage{float}
\usepackage[normalem]{ulem}
\usepackage{makecell}
\usepackage{placeins}
\usepackage[lofdepth,lotdepth,caption=false]{subfig}

\newcommand{\numu}{$\nu_{\mu}$\xspace}

\newcommand{\pizero}{$\pi^{0}$\xspace}

\newcommand{\ncpio}{NC $\pi^0$}

\newcommand{\ncdeltaNgamma}{NC $\Delta$ 1$\gamma$}
\newcommand{\nueCC}{$\nu_{e}$\,CC }
\newcommand{\numuCC}{$\nu_{\mu}$\,CC }

\begin{document}

\preprint{APS/123-QED}

\title{Inclusive Search for Anomalous Single-Photon Production in MicroBooNE}

\include{microboone-author-list-sep2024}

\begin{abstract}
    We present an inclusive search for anomalous production of single-photon events from neutrino interactions in the MicroBooNE experiment. The search and its signal definition are motivated by the previous observation of a low-energy excess of electromagnetic shower events from the MiniBooNE experiment. We use the Wire-Cell reconstruction framework to select a sample of inclusive single-photon final-state interactions with a final efficiency and purity of 7.0\% and 40.2\%, respectively. We leverage simultaneous measurements of sidebands of charged current \numu{} interactions and neutral current interactions producing \pizero{} mesons to constrain signal and background predictions and reduce uncertainties. We perform a blind analysis using a dataset collected from February 2016 to July 2018, corresponding to an exposure of $6.34\times10^{20}$ protons on target from the Booster Neutrino Beam (BNB) at Fermilab. In the full signal region, we observe agreement between the data and the prediction, with a goodness-of-fit $p$-value of 0.11. We then isolate a sub-sample of these events containing no visible protons, and observe $93\pm22\text{(stat.)}\pm35\text{(syst.)}$ data events above prediction, corresponding to just above $2\sigma$ local significance, concentrated at shower energies below 600 MeV.
\end{abstract}

\maketitle

The ``Low Energy Excess'' (LEE) of events with electromagnetic (EM) shower activity reported by the MiniBooNE Collaboration~\cite{mb:2007, mb:2008, mb:2013, mb:2018, mb:2021} is a long-standing anomaly in neutrino physics. It has many proposed explanations, including new types of neutrinos or other physics phenomena beyond the Standard Model (BSM)~\cite{bib:LEEaxion,bib:LEEdarknu,bib:LEEseesaw,bib:LEEpascoli,bib:LEEhiggs,bib:LEEGninenko,bib:LEEkoto,bib:LEEdecay, bib:LEEpseudoscalar, bib:LEEmesondecay}. MiniBooNE is not able to discriminate between electron-induced showers, as expected from the appearance of electron neutrinos ($\nu_e$) from a light sterile neutrino, and events with a single-photon-induced shower in the final state. Therefore, both types of interactions must be examined independently as a source of the LEE.

The MicroBooNE detector is an 85 metric ton active volume liquid argon time projection chamber (LArTPC)~\cite{bib:uB_detector}, situated on-axis with respect to Fermilab’s Booster Neutrino Beam (BNB)~\cite{bib:mbflux} at a distance of 468.5 m from the BNB proton target. This places MicroBooNE only 72.5 m upstream of the MiniBooNE detector hall on the same beamline. The search presented in this Letter uses data corresponding to a BNB exposure of
$6.34 \times 10^{20}$ protons on target (POT), collected from 2016–2018. MicroBooNE's LArTPC technology allows us to distinguish EM showers originating from electrons or photons based on ionization energy deposition at the start of the shower and on the conversion distance of the photon relative to the interaction vertex. 

We report an LEE search result in the inclusive single-photon channel using neutrino-argon scattering data collected by the MicroBooNE experiment. In MicroBooNE’s first round of LEE results, which consisted of both ``electron-like'' searches with electron-initiated EM showers in the final state ~\cite{uB_eLEE_PRL, uB_PeLEE, uB_WCeLEE, uB_DL, pelee15} and ``photon-like'' searches with photon-initiated EM showers in the final state~\cite{uB_gLEE}, no significant excess was observed. Unlike the electron-like LEE searches, which covered a wide range of final states and processes, the photon-like LEE search only focused on a specific Standard Model process --- neutrino-induced neutral current (NC) $\Delta$ radiative decay. In this Letter we expand on our first round of investigations of the MiniBooNE anomaly by presenting a more inclusive search for an excess of photon-like LEE events.

To search for an LEE-like signal we use Monte Carlo (MC) simulations to form event rate predictions for Standard Model-based signal and background processes. A custom tune~\cite{bib:uB_genietune} of the \texttt{GENIE} neutrino event generator software~\cite{Tena-Vidal:2021rpu} is used (\texttt{v3.0.6, G18\_10a\_02\_11a}) to simulate neutrino-argon interactions. We note that this version of \texttt{GENIE} does not simulate some processes relevant to this analysis such as heavier resonances, and coherent single-photon production~\cite{NCCohPaper} -- the contributions from which are expected to be negligible -- and real radiated photons in charged current events~\cite{muonrealrad}.
The BNB flux at MicroBooNE is modeled with the MiniBooNE flux simulation~\cite{bib:mbflux} corrected for MicroBooNE’s beamline position~\cite{bnb_flux}.
Particle transport through the detector is handled by \texttt{GEANT4}~\cite{bib:geant4}, which reproduces ionization and scintillation via dedicated response algorithms ~\cite{bib:uB_signal1, bib:uB_signal2}. Simulated neutrino interactions are overlaid with cosmic-ray data events collected with an unbiased trigger in anti-coincidence with the beam, which allows for data-driven cosmic-ray and detector noise modeling. These tools are implemented using the LArSoft framework~\cite{bib:larsoft}.

The event topology of the MiniBooNE LEE is a single Cherenkov ring consistent with an EM shower (electron or photon). In this photon-like LEE search, we therefore define a signal event as any final state containing one ``reconstructable'' photon-initiated EM shower and any number of charged particles below Cherenkov threshold in MiniBooNE. We define a reconstructable photon-initiated EM shower as one that starts $\geq$3 cm from the TPC boundary and originates from a photon with true energy $\geq$20~MeV. To account for the expected opening angle resolution of our shower reconstruction, we consider two photons (originating from the same vertex) with opening angle $<$20\textdegree{} as one reconstructable photon shower, which is slightly larger than MiniBooNE's quoted resolution of 13\textdegree{}~\cite{mb:2018}. No restrictions are imposed on the number of protons, which are almost always invisible in MiniBooNE due to their high Cherenkov energy threshold of 342 MeV, or charged pions. Similarly, a muon with kinetic energy $<$100~MeV is considered below MiniBooNE's detection threshold~\cite{Miniboonedetector}, permitting them to be included as part of the signal event topology. To probe only photon-like showers, this analysis excludes any events with primary electron-initiated EM showers as signals. 

Based on the final-state topologies described above, we define the following $\nu$-Ar interaction processes modeled in \texttt{GENIE} as our signal:
\begin{enumerate}
    \item NC $\pi^0$ events with only one reconstructable photon shower (\ncpio{} 1$\gamma$),
    \item NC $\Delta$ radiative decay (\ncdeltaNgamma{}),
    \item NC processes that produce a single reconstructable photon from anything other than \pizero{} or $\Delta$ decay, such as $\eta$ or $\rho$ meson decays (NC Other 1$\gamma$), 
    \item \numu{} charged current (\numu{} CC) interactions producing a single reconstructable photon where the muon's kinetic energy is less than 100 MeV (\numuCC{} 1$\gamma$), and
    \item neutrino interactions occurring outside the detector but producing a photon entering and showering inside the TPC fiducial volume (FV), where the FV is defined as a smaller volume 3~cm from the TPC boundary (out of FV 1$\gamma$). \footnote{Due to the differences in size and shape of the MiniBooNE and MicroBooNE detectors, the FV definition is to account for edge effects in MicroBooNE and is not motivated by any MiniBooNE quantities.} 
\end{enumerate}

Due to the existence of category one, \ncpio{} events have the potential to be either signal (if they have one reconstructable photon shower, denoted as NC$\pi^0$ 1$\gamma$) or background (if they have 0 or 2 reconstructable photon showers, denoted as just \ncpio{}). For the final selection, 75\% of the predicted neutrino events contain a \pizero{} decay, regardless of whether they are categorized as signal or background. In addition to Standard Model processes, this signal definition also enables observations of single-photon events induced by possible BSM physics signatures from a broad range of models. By comparing inclusive single-photon data to the Standard Model-based \texttt{GENIE} prediction, we can detect a wide range of physics phenomena that could explain the MiniBooNE LEE.  

We use the Wire-Cell software package~\cite{Qian:2018qbv, Abratenko:2020hpp, wire-cell-pr} for the reconstruction and classification of LArTPC events. The first step of event selection, referred to as pre-selection, is to remove cosmic-ray backgrounds using Wire-Cell's generic neutrino selection. This step identifies the charge and light created by the neutrino interaction within an event and uses this information to reject over $99.99 \%$ of cosmic-ray background events~\cite{generic_nu}. In the process, it also reconstructs the neutrino interaction point, called the neutrino vertex. The reconstructed vertex is required to be more than 5 cm away from the TPC boundary in the drift direction, which removes events interacting near the cathode and anode wire planes that are more likely to be mis-reconstructed. 
We also require at least one reconstructed EM shower to reject a large portion of muon neutrino interaction backgrounds. At this stage, the sample composition is dominated by neutrino interactions with at least one $\pi^0$ meson in the final state and other CC interactions where a shower has been reconstructed.

\begin{figure}[!htbp]
\begin{center}
\includegraphics[trim={0 0 3cm 1cm},clip,width=0.49\textwidth]{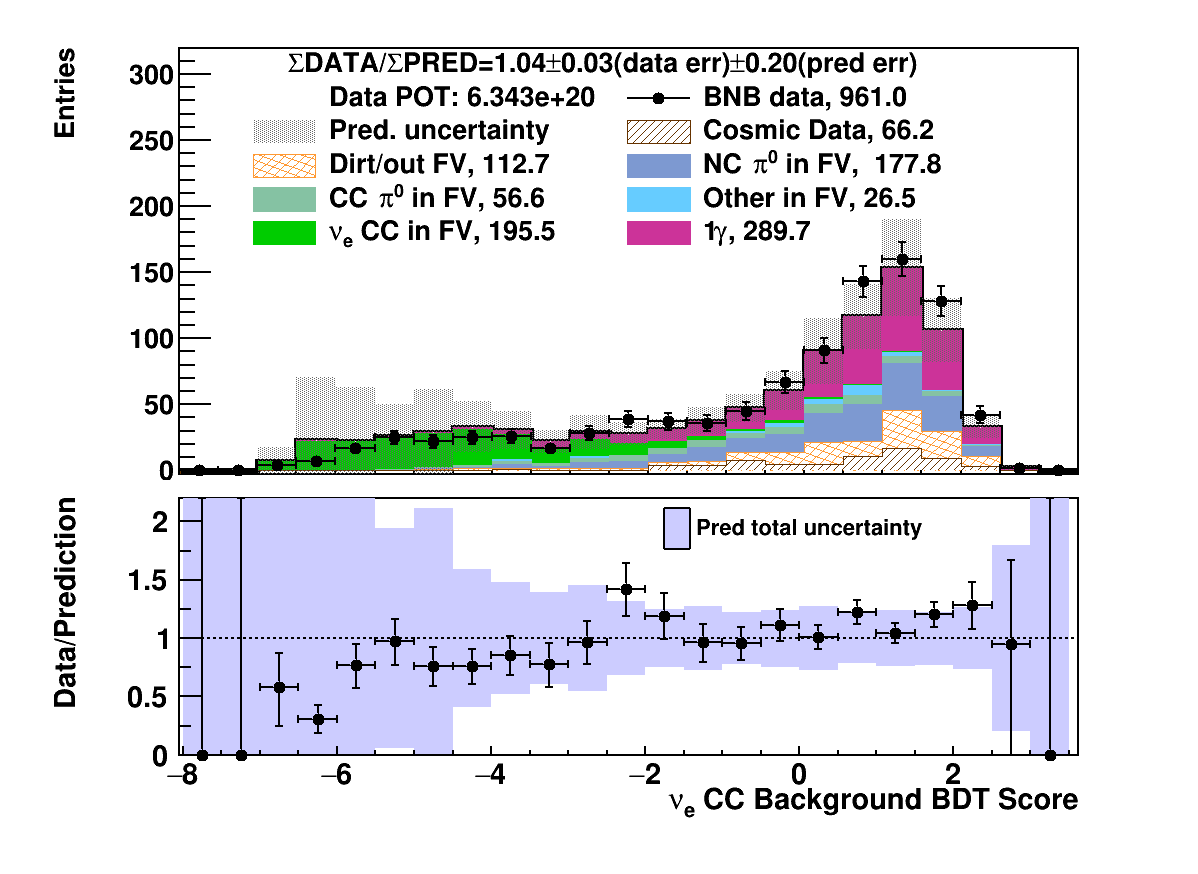}
\put(-180,120){MicroBooNE}
\caption{\label{fig:NUEBDT} Distribution of \nueCC{} background rejection BDT scores. This BDT targets the separation of CC events with a true electron shower (green) from single-photon signal events (pink) by focusing on the calorimetry of the start of the shower ($\mathrm{d}E/\mathrm{d}x$) and shower conversion distances.}
\end{center}
\end{figure}

Following pre-selection, four Boosted Decision Trees (BDTs), utilizing the XGBoost framework~\cite{xgboost}, are used to target specific background categories: $\nu_{\mu}$ CC interactions, NC $\pi^0$ interactions, ``other'' backgrounds (mainly consisting of cosmic rays and neutrinos interacting outside of the TPC), and $\nu_e$ CC interactions. As an example, in the \nueCC{} BDT score distribution (shown in Fig. ~\ref{fig:NUEBDT}), events that contain electron showers (green) and single-photon signal events (pink) are well separated. A BDT score cut is then applied to reject each background category. The cut values are chosen using a simultaneous optimization method which is described in more detail in the Supplemental Material~\cite{bib:supplemental} and~\cite{mythesis}. Events that failed to pass this cut are used as sideband samples for further background studies. Finally, the selection requires exactly one reconstructed EM shower in the event. The ``signal region'' then comprises all events passing these criteria. The final selection achieves a single-photon signal efficiency of $7.0 \%$ with a $40.2 \%$ purity. 

Additionally, a few regions of interest (ROIs) are identified  – before looking at the data in this analysis – to investigate particular topologies that are of interest for the LEE, such as shower energies below 600 MeV and splitting events by proton multiplicities. These regions have been identified and studied in the past by both MiniBooNE~\cite{mb:2021} and MicroBooNE~\cite{uB_gLEE, 2DNCDelPaper, uB_PeLEE}. These ROIs are defined as subsets of the selected inclusive 1$\gamma$ events and are constructed by applying further selection cuts after the nominal ones outlined above.

To validate the modeling of the selected neutrino-induced single-photon events, four sideband samples are defined by inverting selections on the four BDT score distributions. All sidebands show good agreement within systematic uncertainties between data and simulation for kinematic variables used in this analysis, such as shower energy, and for input variables to the BDTs. Examples of these kinematic distributions for the \numuCC{} and \ncpio{} sidebands can be found in the Supplemental Material. The \numuCC{} and \ncpio{} sidebands are used to constrain the background prediction in the signal region in a data-driven approach. Using the framework described in Ref.~\cite{uB_WCeLEE}, the conditional constraint formalism is used to update the central value predictions of the simulation and their respective systematic uncertainties. Comparisons of data to prediction in the sidebands provides constraints to the signal region through correlations of systematic uncertainties in these samples. After constraints, both the background and signal predictions are updated. A notable aspect of the constraint procedure is that it utilizes the correlations between different categories of events. This means that if individual contributions to the CV or systematic errors are constrained separately and added at the end they will not achieve the same result. For this reason, only the total prediction and uncertainty is shown post-constraint.

Systematic uncertainties are estimated on five main aspects of the simulation: the neutrino beam flux, neutrino interactions and particle propagation in the detector, the response of the detector to charge and scintillation light produced by charged particles traversing the liquid argon, Monte Carlo statistical uncertainty, and uncertainty in modeling neutrinos that interact in non-argon material outside of the cryostat, referred to as ``Dirt''. This analysis uses the common systematic framework adopted by MicroBooNE, which varies the central values of underlying parameters that model flux, cross section, and particle re-interaction, independently, within their uncertainties~\cite{bib:mbflux, bib:uB_genietune}. For the modeling of the detector response, uncertainties are evaluated based on the level of agreement between data and simulation in several low-level detector observables~\cite{bib:uB_wiremod}, such as charge signals on the wires and light signals in the PMTs. A Bayesian treatment~\cite{bayes} is used to account for MC statistical uncertainties. The Dirt uncertainty is an additional, relative 50\% bin-to-bin uncorrelated uncertainty arising from the modeling of the materials outside the cryostat. Systematic uncertainties are incorporated through a covariance matrix.

Table~\ref{tab:uncert} shows the systematic uncertainty on the total number of predicted events in the signal region. The dominant uncertainty is associated with cross section modeling of $\nu$-Ar interactions, particularly the modeling of resonant interactions, which is the largest background category due to the mis-identification rate for $\pi^0$ final states reconstructed as single-photon events. 

\begin{table}[!htbp]
  \centering
  \begin{tabular}{  l  c  } 
      \hline\hline
       Type of Uncertainty \hspace{37mm} & Selection \\ 
      \hline
      
      Flux model & $6.4 \%$ \\
     
      Cross section and Re-interaction models & $19.1 \%$ \\
       
      Detector response & $6.5 \%$ \\
       
      MC statistics & $2.0 \%$ \\
       
      Interactions in Dirt & $0.8 \%$ \\
       \hline
      Total Uncertainty (Unconstrained) & $21.3 \%$ \\
       
      Total Uncertainty (Constrained) & $8.4 \%$ \\
      \hline\hline
  \end{tabular}
   \caption{Uncertainty on the number of events after the inclusive single-photon selection. Numbers are derived by considering an integrated shower energy between 0 and 1500 MeV.} 
   \label{tab:uncert}
\end{table}

This analysis adheres to a signal-blind analysis strategy, whereby the data in the signal region are kept blinded until the analysis procedure has been fully developed. Unblinding occurs after all background modeling validations have been completed, the selection frozen, and several fake-data studies on the signal region performed. Upon unblinding the single-photon signal region, we observed $678$ data events with an expected constrained prediction of $564 \pm 24\text{(stat.)} \pm 51$(syst.).

The numbers of final selected events from prediction and data are summarized in Table~\ref{tab:Xpbreakdown}. For the prediction, the inclusive 1$\gamma$ signal consists of five SM processes as previously described. The breakdown of the total 1$\gamma$ signal into these categories is shown.
\ncpio{} interactions inside the FV with two visible photons make up 20\% of all selected events, making them the largest background contribution in the final selection.

\begin{table}[!htbp]
        \centering
        \begin{tabular}{ l c  c } 
            \hline
            \hline
            Process \hspace{30mm} & Selected Events \\ 
            \hline
            Total Signal: 1$\gamma$ & $247 \pm 69$ \\
            \hspace{5mm} \ncdeltaNgamma{}    & \hspace{5mm} $19 \pm 21 $  \\
            \hspace{5mm} \ncpio{} 1$\gamma$  & \hspace{5mm} $143 \pm 51 $ \\
            \hspace{5mm} NC Other 1$\gamma$  & \hspace{5mm} $3 \pm 3 $    \\
            \hspace{5mm} \numuCC{} 1$\gamma$ & \hspace{5mm} $9 \pm 2 $    \\
            \hspace{5mm} Out of FV 1$\gamma$ & \hspace{5mm} $73 \pm 17 $  \\
            \hline
            Cosmic data \hspace{15mm}  & $57 \pm 7$  \\
            Dirt/out FV \hspace{15mm}  & $101 \pm 25$ \\
            Neutrino backgrounds in FV & $204 \pm 45$  \\
            \hline
            Total prediction (unconstr.) & \hspace{1mm} $608 \pm 25 \text{(stat.)} \pm 128 \text{(syst.)}$ \\
            Total prediction (constr.) & $564 \pm 24 \text{(stat.)} \pm 51 \text{(syst.)}$ \\
            \hline
            BNB data & $678$ \\
            \hline
            \hline
        \end{tabular}
         \caption{Event number in the signal region by process for the inclusive single-photon final selection, using $6.34 \times 10^{20}$ POT of BNB data.}
         \label{tab:Xpbreakdown}
\end{table}

\begin{figure}[!htbp]
\begin{center}
    \subfloat[]{%
        \includegraphics[trim={1.25cm 0.15cm 0.1cm 0.5cm},clip,width=0.49\textwidth]{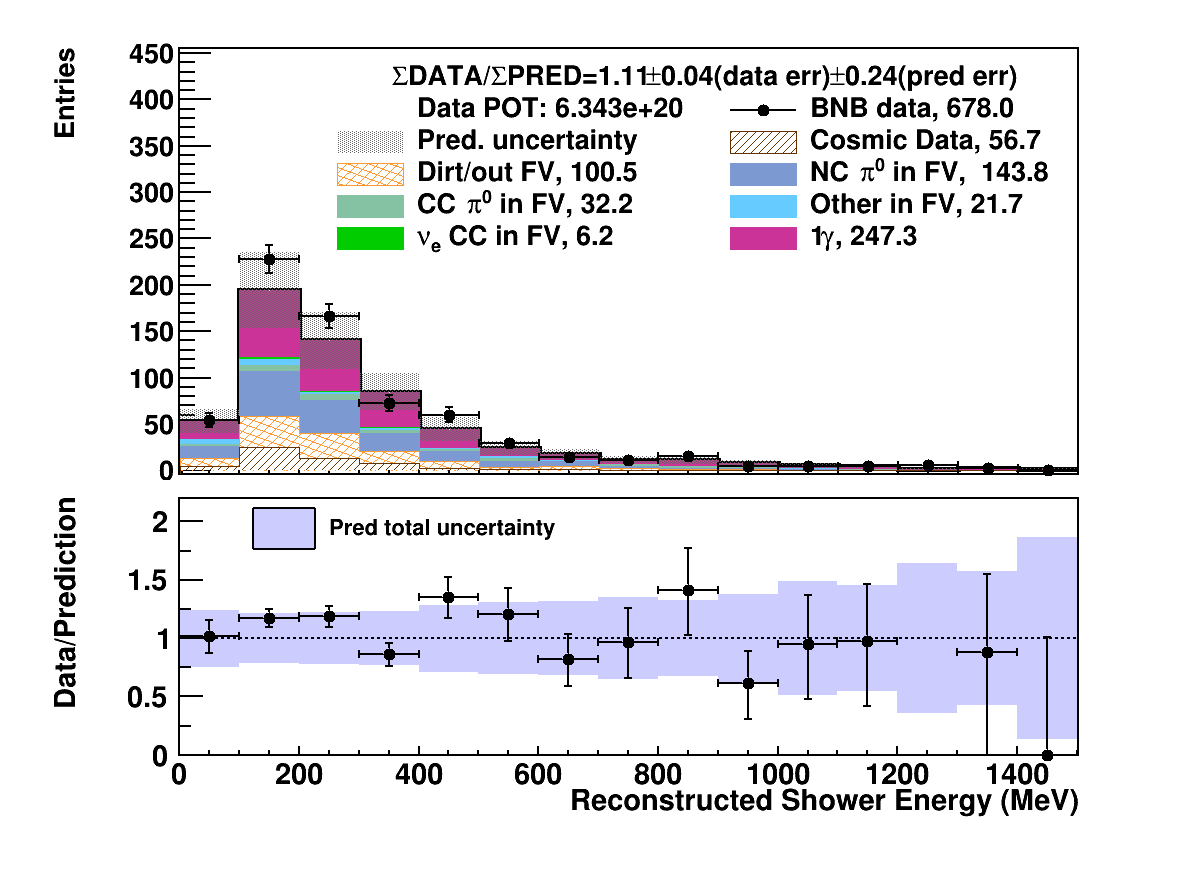}\label{fig:results_showerenergy_fullrange}
        \put(-90,115){MicroBooNE}
        \put(-90,105){unconstrained}%
    }\\
    \subfloat[]{%
        \includegraphics[trim={1.25cm 0.15cm 0.15cm 0.5cm},clip,width=0.49\textwidth]{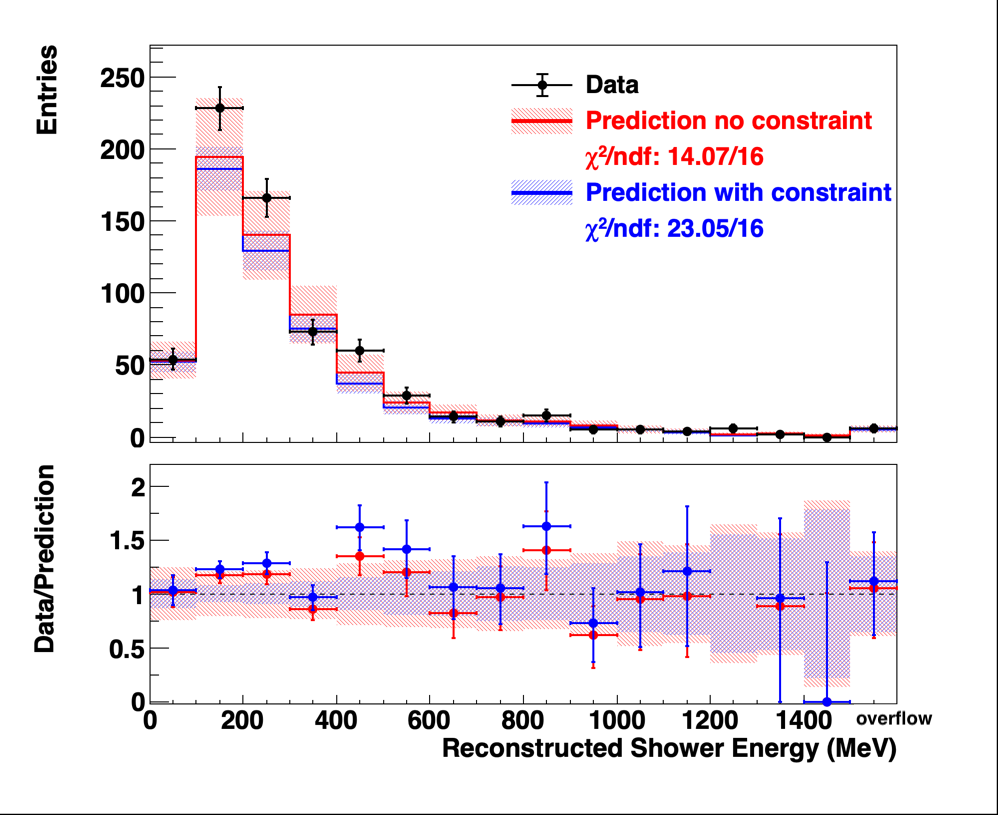}\label{fig:results_showerenergy_fullrange_constrained}
        \put(-90,115){MicroBooNE}%
    }
    \caption{ Reconstructed shower energy for inclusive single-photon selected events, (a) before constraint and (b) before (red) and after (blue) constraint. (a) shows the inclusive single-photon signal events in pink. The horizontal axis is reconstructed shower energy in 100 MeV bins. Events with energy above 1500 MeV are included via an overflow bin. The constrained prediction has been pulled down in many bins, due to the \ncpio{} constraining channel predicting more events than seen in the data.}\label{fig:results_showerenergy}
\end{center}
\end{figure}

The shower energy distribution of final selected events, including 1$\gamma$ signal processes and predicted background processes broken down into 6 categories, is shown before constraint in Fig.~\ref{fig:results_showerenergy_fullrange}. A comparison of the data to the prediction after sideband constraints is shown in  Fig.~\ref{fig:results_showerenergy_fullrange_constrained}, where the uncertainty is significantly reduced and the central value is pulled down, mainly due to an over-prediction in the \ncpio{} sideband as shown in the Supplemental Material. The shower energy spectra are consistent between data and simulation throughout the entire energy range, with a $\chi^2$/n.d.f. after constraint of 23/16, corresponding to a $p$-value of $0.11$. In the region below 600 MeV, a mild data excess with a $p$-value of 0.03 is observed.

\begin{figure}[!htbp]
\begin{center}
\includegraphics[trim={0 0.15cm 0.15cm 0},clip,width=0.49\textwidth]{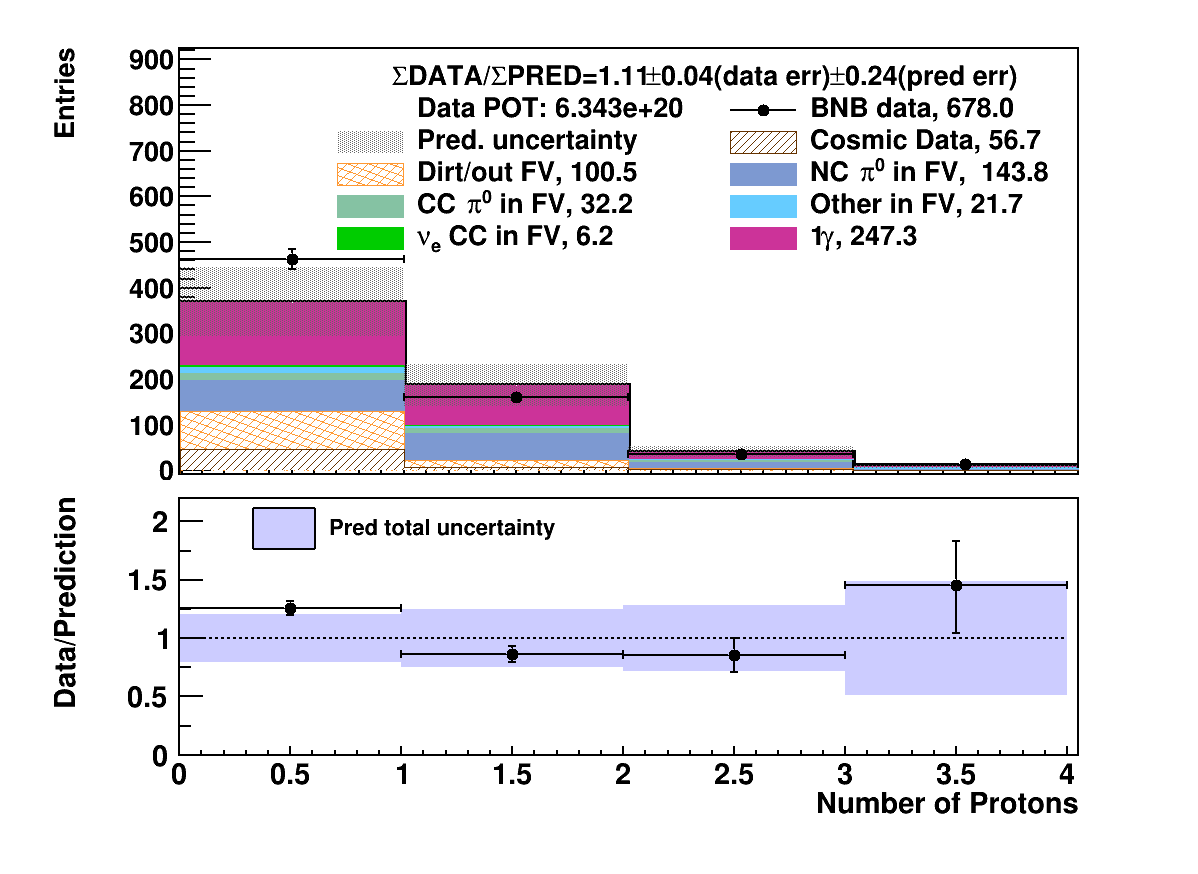}
\put(-90,112){MicroBooNE}
\put(-90,102){unconstrained}
\caption{\label{fig:results_numproton_fullrange_constrained} Numbers of reconstructed primary protons for inclusive single-photon selected events with no constraint applied. The horizontal axis is number of reconstructed protons with kinetic energy above the 35 MeV threshold.} 
\end{center}
\end{figure}

One advantage of MicroBooNE's LArTPC detector compared to MiniBooNE's oil Cherenkov detector is the ability to detect protons. Prior analysis~\cite{bib:ben_numuCC} has shown that Wire-Cell particle identification can reliably reconstruct the physics attributes of protons with true kinetic energies as low as 35 MeV. While LArTPCs have demonstrated the potential to perform proton reconstruction below this energy range~\cite{blipreco, NCCohPaper}, this analysis focuses on and quotes multiplicities of protons only above this 35 MeV threshold. Fig.~\ref{fig:results_numproton_fullrange_constrained} shows the numbers of reconstructed protons above threshold in the selected inclusive 1$\gamma$ sample. We observe a data excess in the zero proton bin, while events showing one or more protons show good agreement. 

To further investigate the nature of the discrepancy, we perform tests using the previously identified ROIs. We define one ROI as the subset of selected inclusive 1$\gamma$ events that contain no reconstructed protons in the final state (1$\gamma$0p) and have reconstructed shower energy $\leq 600$~MeV. We use the same \numuCC{} and \ncpio{} sidebands to constrain this ROI, but with the zero-proton requirement also applied in order to reduce the effect of proton counting errors when correlating the sideband and signal samples in the constraining procedure. The resulting selection achieves an efficiency and purity of true single-photon events with no protons 
of $10 \%$ and $24 \%$, respectively. Note that since the proton threshold is 35~MeV, it is possible for the events in this sample to contain a low energy proton. Our prediction shows about 10\% of the selected $1\gamma 0p$ events contain a proton with kinetic energy below 35 MeV.

\begin{figure}[!htbp]
\begin{center}
\includegraphics[trim={1.0cm 0.18cm 0.2cm 0.5cm},clip,width=0.49\textwidth]{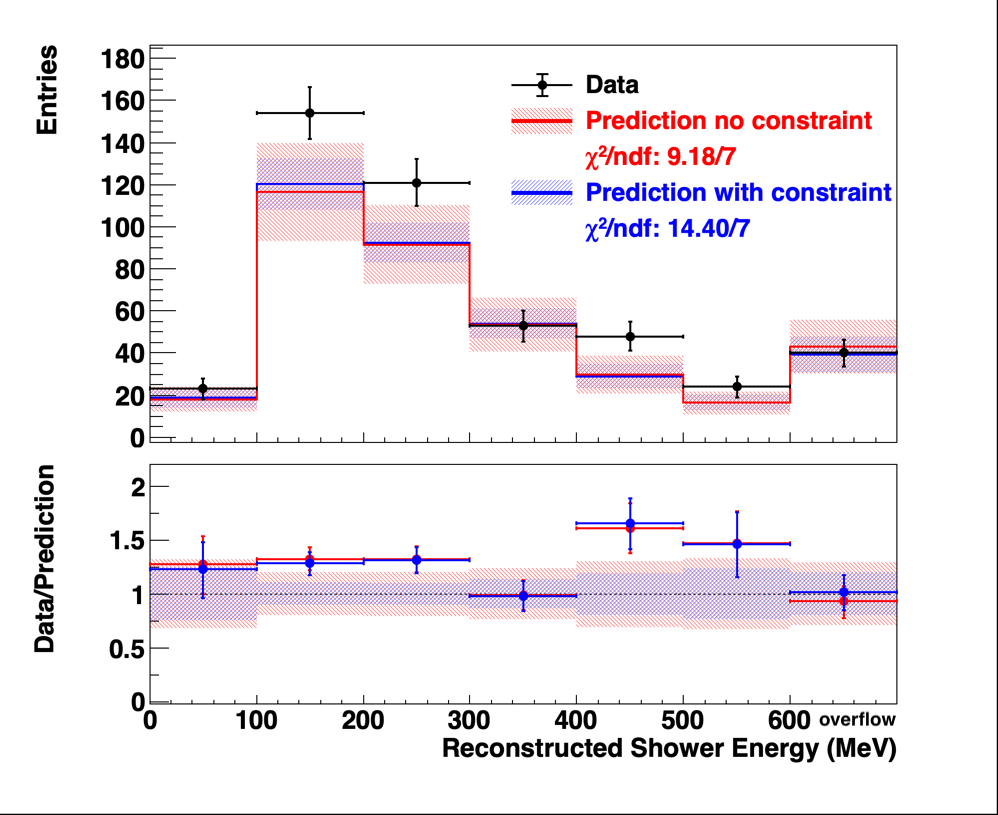}
\put(-185, 205){single-photon, zero-proton events}
\put(-90,137){MicroBooNE}
\caption{\label{fig:results_showerenergy_0p_ROI_constrained} Reconstructed shower energy for inclusive single-photon, zero-proton selected events, before and after constraint. The horizontal axis is reconstructed shower energy in 100 MeV bins with the last bin containing overflow events.}
\end{center}
\end{figure}

The reconstructed shower energy for this ROI sample is shown in Fig.~\ref{fig:results_showerenergy_0p_ROI_constrained}. The central value slightly increases after constraints for the 1$\gamma$0p sample, which is due to the 4\% data excess in the zero proton \ncpio{} sideband. The constraint's impact on the central value for 0p is opposite to the nominal (Xp, X~$\geq 0$) sample, due to a $26 \%$ data deficit in the Np (N~$ > 0$) sample of the \ncpio{} sideband. The 0p and Np distributions of these sidebands can be found in the Supplemental Material, along with more kinematic distributions of the excess events. For Fig.~\ref{fig:results_showerenergy_0p_ROI_constrained}, we analyze the goodness-of-fit using a $\chi^2$ distribution consisting of 4 million pseudo-experiments. The local significance of the excess in the ROI is 2.2$\sigma$. Since this ROI was identified as of interest before unblinding, no look elsewhere effect evaluation was performed.

\begin{figure}[!htbp]
\begin{center}
\includegraphics[width=0.49\textwidth]{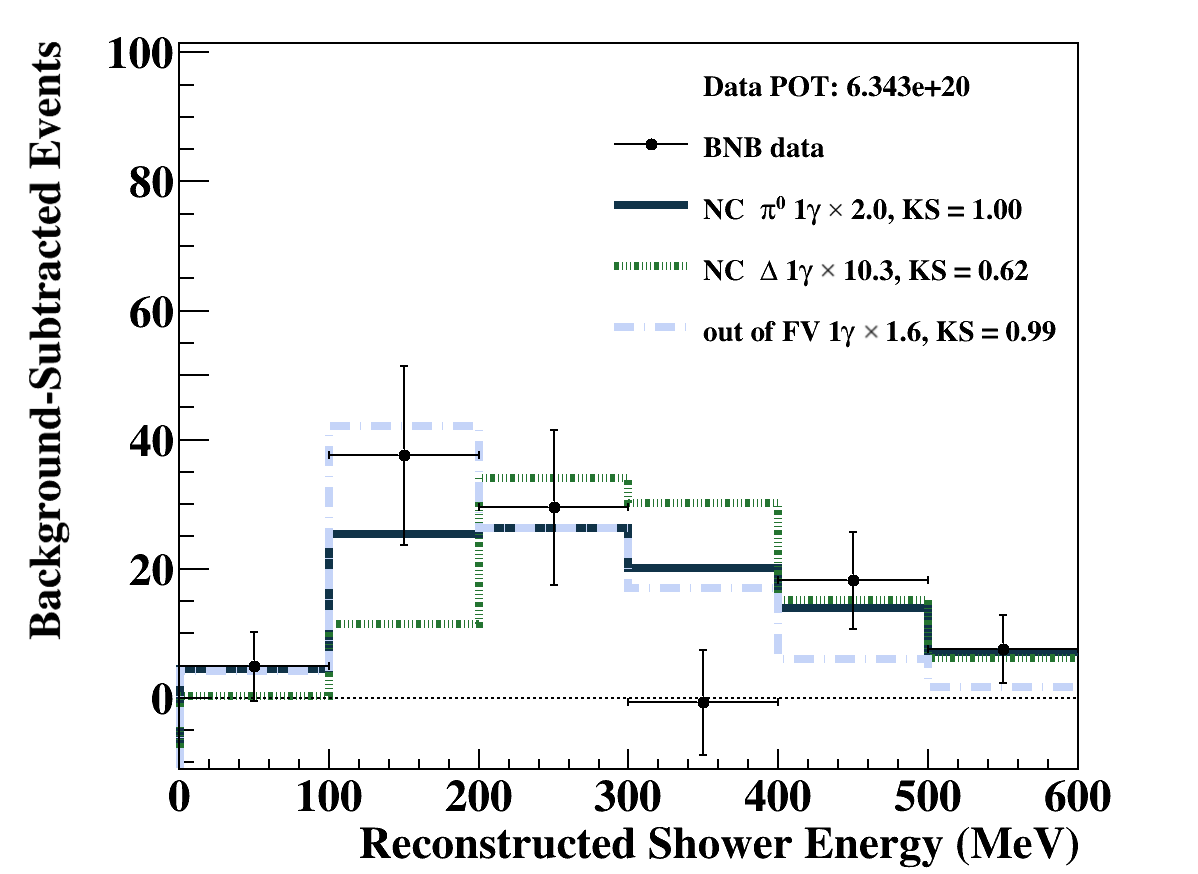}
\put(-190,152){MicroBooNE}
\caption{\label{fig:results_3template_0p_ROI} Shape-only fit for $1\gamma$0p background-subtracted events (data minus constrained prediction). Three signal categories are shown alongside the best fit scaling parameter and the Kolmogorov–Smirnov test statistic. No constraint or systematic uncertainties on the shape of the individual predicted signal categories are considered. The error bars on the data points include both statistical and constrained systematic errors.}
\end{center}
\end{figure}

In a separate study, we test the compatibility of the background-subtracted (``excess'') events with various signal processes in a shape-only comparison of the reconstructed shower energy distribution in the ROI. Each of the five signal categories is scaled up to match the integrated event excess, neglecting any correlations with background and sideband predictions. The normalization factors and Kolmogorov–Smirnov (KS) test statistics calculated with constrained systematic uncertainties are reported to quantify the comparison. Fig.~\ref{fig:results_3template_0p_ROI} shows the results for three representative signal processes.  
The shape of \ncdeltaNgamma{} events peaks at higher energies than the excess events and requires a scaling factor of 10.3, which is ruled out with high significance by previous MicroBooNE results~\cite{uB_gLEE, 2DNCDelPaper}. In contrast, scaling the normalization of \ncpio{} $1\gamma$ in FV events by a factor of 2 and out of FV $1\gamma$ events (of which 76\% come from \ncpio{} interactions) by 1.6 both show good agreement with the shape of the excess. This suggests that part of the observed excess events could originate from either or both of these processes, which can enter the signal sample due to the relatively large conversion distance for photons combined with the elongated shape of the MicroBooNE detector.

The final study we conduct is to compare the observed excess to a photon interpretation of the MiniBooNE LEE, referred to as the ``LEE-$\gamma$ model''. 
A detailed description of this model can be found in the Supplemental Material. When scaled from MiniBooNE to MicroBooNE under a target nucleon interaction assumption, the LEE-$\gamma$ model predicts a 33 event excess in the $1\gamma0p$ sample, which is less than the observed excess of $93\pm22\text{(stat.)}\pm35\text{(syst.)}$ events.
 A $\Delta \chi^2$ test statistic constructed using the combined Neyman-Pearson (CNP)~\cite{CNP} method is used to simultaneously compare the $1\gamma0p$ data sample to the constrained, nominal \texttt{GENIE} prediction ($H_0$) and constrained \texttt{GENIE} plus LEE-$\gamma$ model ($H_1$). Our data shows agreement with $H_1$ with a $p$-value of 0.14 and with $H_0$ with a $p$-value of 0.02 in the shower energy distribution. Detailed analyses of the goodness-of-fit in both shower energy and angle distributions are shown in the Supplementary Material.

In summary, we present the first result of MicroBooNE's LEE search in the inclusive single-photon channel using $6.34 \times 10^{20}$ POT of data collected from Fermilab's Booster Neutrino Beam. We select $678$ data events, corresponding to a $20 \%$ data excess compared to the prediction of $564 \pm 24\text{(stat.)} \pm 51\text{(syst.)}$ events. In the full range of reconstructed shower energy and after a data-driven constraint procedure leveraging \numuCC{} and \ncpio{} rich sidebands is applied, the data spectrum is consistent with the prediction with a $p$-value of 0.11. While further investigation shows that the shape of some shower kinematic distributions are potentially compatible with mismodeled out of FV and \ncpio{} events, the required scaling factors for \ncpio{} events would push the predictions well out of their uncertainties in the relevant background-rich sidebands. We, therefore, have so far not identified a simple and complete explanation for the observed excess. The excess events are concentrated in the phase space with shower energy below 600 MeV and no detectable protons in the final state. This region of phase-space shows a 2.2$\sigma$ local significance compared to the constrained \texttt{GENIE} prediction. This result presents a first look at the MiniBooNE LEE in an inclusive photon channel, and motivates further investigations both with MicroBooNE's full dataset, which contains 1.3 $\times$ $10^{21}$ POT of data, as well as within the broader SBN program~\cite{bib:SBN}. A data release corresponding to this analysis is available at Ref.~\cite{inclusive_hepdata}.

\FloatBarrier{}

This document was prepared by the MicroBooNE collaboration using the
resources of the Fermi National Accelerator Laboratory (Fermilab), a
U.S. Department of Energy, Office of Science, Office of High Energy Physics HEP User Facility.
Fermilab is managed by Fermi Forward Discovery Group, LLC, acting
under Contract No. 89243024CSC000002.  MicroBooNE is supported by the
following: 
the U.S. Department of Energy, Office of Science, Offices of High Energy Physics and Nuclear Physics; 
the U.S. National Science Foundation; 
the Swiss National Science Foundation; 
the Science and Technology Facilities Council (STFC), part of the United Kingdom Research and Innovation; 
the Royal Society (United Kingdom); 
the UK Research and Innovation (UKRI) Future Leaders Fellowship; 
and the NSF AI Institute for Artificial Intelligence and Fundamental Interactions. 
Additional support for 
the laser calibration system and cosmic ray tagger was provided by the 
Albert Einstein Center for Fundamental Physics, Bern, Switzerland. We 
also acknowledge the contributions of technical and scientific staff 
to the design, construction, and operation of the MicroBooNE detector 
as well as the contributions of past collaborators to the development 
of MicroBooNE analyses, without whom this work would not have been 
possible. 
For the purpose of open access, the authors have applied 
a Creative Commons Attribution (CC BY) public copyright license to 
any Author Accepted Manuscript version arising from this submission.

\bibliography{zapsamp}

\end{document}

%% file: microboone-author-list-sep2024.tex
\newcommand{\ANL}{Argonne National Laboratory (ANL), Lemont, IL, 60439, USA}
\newcommand{\Bern}{Universit{\"a}t Bern, Bern CH-3012, Switzerland}
\newcommand{\BNL}{Brookhaven National Laboratory (BNL), Upton, NY, 11973, USA}
\newcommand{\UCSB}{University of California, Santa Barbara, CA, 93106, USA}
\newcommand{\Cambridge}{University of Cambridge, Cambridge CB3 0HE, United Kingdom}
\newcommand{\CIEMAT}{Centro de Investigaciones Energ\'{e}ticas, Medioambientales y Tecnol\'{o}gicas (CIEMAT), Madrid E-28040, Spain}
\newcommand{\Chicago}{University of Chicago, Chicago, IL, 60637, USA}
\newcommand{\Cincinnati}{University of Cincinnati, Cincinnati, OH, 45221, USA}
\newcommand{\CSU}{Colorado State University, Fort Collins, CO, 80523, USA}
\newcommand{\Columbia}{Columbia University, New York, NY, 10027, USA}
\newcommand{\Edinburgh}{University of Edinburgh, Edinburgh EH9 3FD, United Kingdom}
\newcommand{\FNAL}{Fermi National Accelerator Laboratory (FNAL), Batavia, IL 60510, USA}
\newcommand{\Granada}{Universidad de Granada, Granada E-18071, Spain}
\newcommand{\IIT}{Illinois Institute of Technology (IIT), Chicago, IL 60616, USA}
\newcommand{\ICL}{Imperial College London, London SW7 2AZ, United Kingdom}
\newcommand{\Indiana}{Indiana University, Bloomington, IN 47405, USA}
\newcommand{\KSU}{Kansas State University (KSU), Manhattan, KS, 66506, USA}
\newcommand{\Lancaster}{Lancaster University, Lancaster LA1 4YW, United Kingdom}
\newcommand{\LANL}{Los Alamos National Laboratory (LANL), Los Alamos, NM, 87545, USA}
\newcommand{\Louisiana}{Louisiana State University, Baton Rouge, LA, 70803, USA}
\newcommand{\Manchester}{The University of Manchester, Manchester M13 9PL, United Kingdom}
\newcommand{\MIT}{Massachusetts Institute of Technology (MIT), Cambridge, MA, 02139, USA}
\newcommand{\Michigan}{University of Michigan, Ann Arbor, MI, 48109, USA}
\newcommand{\MSU}{Michigan State University, East Lansing, MI 48824, USA}
\newcommand{\Minnesota}{University of Minnesota, Minneapolis, MN, 55455, USA}
\newcommand{\Nankai}{Nankai University, Nankai District, Tianjin 300071, China}
\newcommand{\NMSU}{New Mexico State University (NMSU), Las Cruces, NM, 88003, USA}
\newcommand{\Oxford}{University of Oxford, Oxford OX1 3RH, United Kingdom}
\newcommand{\Pitt}{University of Pittsburgh, Pittsburgh, PA, 15260, USA}
\newcommand{\QMUL}{Queen Mary University of London, London E1 4NS, United Kingdom}
\newcommand{\Rutgers}{Rutgers University, Piscataway, NJ, 08854, USA}
\newcommand{\SLAC}{SLAC National Accelerator Laboratory, Menlo Park, CA, 94025, USA}
\newcommand{\SDSMT}{South Dakota School of Mines and Technology (SDSMT), Rapid City, SD, 57701, USA}
\newcommand{\Maine}{University of Southern Maine, Portland, ME, 04104, USA}
\newcommand{\Syracuse}{Syracuse University, Syracuse, NY, 13244, USA}
\newcommand{\TelAviv}{Tel Aviv University, Tel Aviv, Israel, 69978}
\newcommand{\UTA}{University of Texas, Arlington, TX, 76019, USA}
\newcommand{\Tufts}{Tufts University, Medford, MA, 02155, USA}
\newcommand{\VTech}{Center for Neutrino Physics, Virginia Tech, Blacksburg, VA, 24061, USA}
\newcommand{\Warwick}{University of Warwick, Coventry CV4 7AL, United Kingdom}

\affiliation{\ANL}
\affiliation{\Bern}
\affiliation{\BNL}
\affiliation{\UCSB}
\affiliation{\Cambridge}
\affiliation{\CIEMAT}
\affiliation{\Chicago}
\affiliation{\Cincinnati}
\affiliation{\CSU}
\affiliation{\Columbia}
\affiliation{\Edinburgh}
\affiliation{\FNAL}
\affiliation{\Granada}
\affiliation{\IIT}
\affiliation{\ICL}
\affiliation{\Indiana}
\affiliation{\KSU}
\affiliation{\Lancaster}
\affiliation{\LANL}
\affiliation{\Louisiana}
\affiliation{\Manchester}
\affiliation{\MIT}
\affiliation{\Michigan}
\affiliation{\MSU}
\affiliation{\Minnesota}
\affiliation{\Nankai}
\affiliation{\NMSU}
\affiliation{\Oxford}
\affiliation{\Pitt}
\affiliation{\QMUL}
\affiliation{\Rutgers}
\affiliation{\SLAC}
\affiliation{\SDSMT}
\affiliation{\Maine}
\affiliation{\Syracuse}
\affiliation{\TelAviv}
\affiliation{\UTA}
\affiliation{\Tufts}
\affiliation{\VTech}
\affiliation{\Warwick}

\author{P.~Abratenko} \affiliation{\Tufts}
\author{D.~Andrade~Aldana} \affiliation{\IIT}
\author{L.~Arellano} \affiliation{\Manchester}
\author{J.~Asaadi} \affiliation{\UTA}
\author{A.~Ashkenazi}\affiliation{\TelAviv}
\author{S.~Balasubramanian}\affiliation{\FNAL}
\author{B.~Baller} \affiliation{\FNAL}
\author{A.~Barnard} \affiliation{\Oxford}
\author{G.~Barr} \affiliation{\Oxford}
\author{D.~Barrow} \affiliation{\Oxford}
% ...

\collaboration{The MicroBooNE Collaboration}
\thanks{microboone\_info@fnal.gov}\noaffiliation

%% file: zapsamp.bib
@PREAMBLE{
 "\providecommand{\noopsort}[1]{}" 
 # "\providecommand{\singleletter}[1]{#1}%" 
}

@misc{bib:supplemental,
    note = "See Supplemental Material at \url{url placeholder} for information about systematic errors and efficiencies, BDT score distributions, distributions and studies for the constraining channels, information on the LEE-$\gamma$ model, kinematic distributions for single-photon events with 1 or more protons, and an event display for a signal event."
}

@article{bib:LEEdecay,
    author = "Dentler, Mona and Esteban, Ivan and Kopp, Joachim and Machado, Pedro",
    title = "{Decaying Sterile Neutrinos and the Short Baseline Oscillation Anomalies}",
    reportNumber = "FERMILAB-PUB-19-538-T",
    doi = "10.1103/PhysRevD.101.115013",
    journal = "Phys. Rev. D",
    volume = "101",
    number = "11",
    pages = "115013",
    year = "2020"
}

@article{bib:LEEkoto,
    author = "Dutta, Bhaskar and Ghosh, Sumit and Li, Tianjun",
    title = "{Explaining $(g-2)_{\mu,e}$, the KOTO anomaly and the MiniBooNE excess in an extended Higgs model with sterile neutrinos}",
    reportNumber = "MI-TH-2012",
    doi = "10.1103/PhysRevD.102.055017",
    journal = "Phys. Rev. D",
    volume = "102",
    number = "5",
    pages = "055017",
    year = "2020"
}

@article{bib:LEEGninenko,
    author = "Gninenko, Sergei N.",
    title = "{A resolution of puzzles from the LSND, KARMEN, and MiniBooNE experiments}",
    doi = "10.1103/PhysRevD.83.015015",
    journal = "Phys. Rev. D",
    volume = "83",
    pages = "015015",
    year = "2011"
}

@article{bib:LEEhiggs,
    author = "Abdallah, Waleed and Gandhi, Raj and Roy, Samiran",
    title = "{Two-Higgs doublet solution to the LSND, MiniBooNE and muon g-2 anomalies}",
    doi = "10.1103/PhysRevD.104.055028",
    journal = "Phys. Rev. D",
    volume = "104",
    number = "5",
    pages = "055028",
    year = "2021"
}

@article{bib:LEEpseudoscalar,
      title = "{A 17 MeV pseudoscalar and the LSND, MiniBooNE and ATOMKI anomalies}", 
      author={Waleed Abdallah and Raj Gandhi and Tathgata Ghosh and Najimuddin Khan and Samiran Roy and Subhojit Roy},
      year={2024},
      eprint={2406.07643},
      archivePrefix={arXiv},
      primaryClass={hep-ph},
      url={https://arxiv.org/abs/2406.07643}, 
      journal = {arXiv e-prints}
}

@article{bib:LEEmesondecay,
  title = "{Solutions to the MiniBooNE Anomaly from New Physics in Charged Meson Decays}",
  author = {Dutta, Bhaskar and Kim, Doojin and Thompson, Adrian and Thornton, Remington T. and Van de Water, Richard G.},
  journal = {Phys. Rev. Lett.},
  volume = {129},
  issue = {11},
  pages = {111803},
  numpages = {7},
  year = {2022},
  month = {Sep},
  publisher = {American Physical Society},
  doi = {10.1103/PhysRevLett.129.111803},
  url = {https://link.aps.org/doi/10.1103/PhysRevLett.129.111803}
}

@article{bib:LEEpascoli,
    author = "Ballett, Peter and Pascoli, Silvia and Ross-Lonergan, Mark",
    title = "{U(1)' mediated decays of heavy sterile neutrinos in MiniBooNE}",
    reportNumber = "IPPP/18/70",
    doi = "10.1103/PhysRevD.99.071701",
    journal = "Phys. Rev. D",
    volume = "99",
    pages = "071701",
    year = "2019"
}

@article{bib:LEEseesaw,
    author = "Abdullahi, Asli and Hostert, Matheus and Pascoli, Silvia",
    title = "{A dark seesaw solution to low energy anomalies: MiniBooNE, the muon (g\ensuremath{-}2), and BaBar}",
    reportNumber = "FTPI-MINN-20-25, IPPP/20/32, FTPI-MINN-20-25,IPPP/20/32",
    doi = "10.1016/j.physletb.2021.136531",
    journal = "Phys. Lett. B",
    volume = "820",
    pages = "136531",
    year = "2021"
}

@article{bib:LEEdarknu,
    author = "Bertuzzo, Enrico and Jana, Sudip and Machado, Pedro A. N. and Zukanovich Funchal, Renata",
    title = "{Dark Neutrino Portal to Explain MiniBooNE excess}",
    reportNumber = "FERMILAB-PUB-18-336-T, OSU-HEP-18-04",
    doi = "10.1103/PhysRevLett.121.241801",
    journal = "Phys. Rev. Lett.",
    volume = "121",
    number = "24",
    pages = "241801",
    year = "2018"
}

@article{bib:LEEaxion,
    author = "Chang, Chia-Hung Vincent and Chen, Chuan-Ren and Ho, Shu-Yu and Tseng, Shih-Yen",
    title = "{Explaining the MiniBooNE anomalous excess via a leptophilic ALP-sterile neutrino coupling}",
    reportNumber = "KIAS-P21006",
    doi = "10.1103/PhysRevD.104.015030",
    journal = "Phys. Rev. D",
    volume = "104",
    number = "1",
    pages = "015030",
    year = "2021"
}

@article{bib:SBN,
    author = "Antonello, M. and others",
    collaboration = "MicroBooNE, LAr1-ND, ICARUS-WA104",
    title = "{A Proposal for a Three Detector Short-Baseline Neutrino Oscillation Program in the Fermilab Booster Neutrino Beam}",
    eprint = "1503.01520",
    archivePrefix = "arXiv",
    primaryClass = "physics.ins-det",
    month = "3",
    year = "2015",
    journal = {arXiv e-prints}
    
}

@article{bib:geant4,
    author = "Agostinelli, S. and others",
    collaboration = "GEANT4",
    title = "{GEANT4--a simulation toolkit}",
    reportNumber = "SLAC-PUB-9350, FERMILAB-PUB-03-339, CERN-IT-2002-003",
    doi = "10.1016/S0168-9002(03)01368-8",
    journal = "Nucl. Instrum. Meth. A",
    volume = "506",
    pages = "250--303",
    year = "2003"
}

@article{bib:uB_detector,
    author = "Acciarri, R. and others",
    collaboration = "MicroBooNE Collaboration",
    title = "{Design and Construction of the MicroBooNE Detector}",
    reportNumber = "FERMILAB-PUB-16-613-ND",
    doi = "10.1088/1748-0221/12/02/P02017",
    journal = "J. Instrum{.}",
    volume = "12",
    number = "02",
    pages = {P02017},
    year = {2017}
}

@article{bib:larsoft,
    author = "Snider, E. L. and Petrillo, G.",
    editor = "Mount, Richard and Tull, Craig",
    title = "{LArSoft: Toolkit for Simulation, Reconstruction and Analysis of Liquid Argon TPC Neutrino Detectors}",
    reportNumber = "FERMILAB-CONF-17-052-CD",
    doi = "10.1088/1742-6596/898/4/042057",
    journal = "J. Phys. Conf. Ser.",
    volume = "898",
    number = "4",
    pages = "042057",
    year = "2017"
}

@article{bib:mbflux,
    author = "Aguilar-Arevalo, A. A. and others",
    collaboration = "MiniBooNE Collaboration",
    title = "{The Neutrino Flux prediction at MiniBooNE}",
    doi = "10.1103/PhysRevD.79.072002",
    journal = "Phys. Rev. D",
    volume = "79",
    pages = "072002",
    year = "2009"
}

@article{bib:uB_genietune,
    author = "Abratenko, P. and others",
    collaboration = "MicroBooNE Collaboration",
    title = "{New $CC0\pi$ GENIE model tune for MicroBooNE}",
    reportNumber = "FERMILAB-PUB-21-508-AD-ND-SCD-TD",
    doi = "10.1103/PhysRevD.105.072001",
    journal = "Phys. Rev. D",
    volume = "105",
    number = "7",
    pages = "072001",
    year = "2022"
}

@article{bib:uB_wiremod,
    author = "Abratenko, P. and others",
    collaboration = "MicroBooNE Collaboration",
    title = "{Novel approach for evaluating detector-related uncertainties in a LArTPC using MicroBooNE data}",
    reportNumber = "FERMILAB-PUB-21-501-ND-SCD",
    doi = "10.1140/epjc/s10052-022-10270-8",
    journal = "Eur. Phys. J. C",
    volume = "82",
    number = "5",
    pages = "454",
    year = "2022"
}

@article{bib:uB_signal1,
    author = "Adams, C. and others",
    collaboration = "MicroBooNE Collaboration",
    title = "{Ionization electron signal processing in single phase LArTPCs. Part I. Algorithm Description and quantitative evaluation with MicroBooNE simulation}",
    reportNumber = "FERMILAB-PUB-18-056-ND-PPD",
    doi = "10.1088/1748-0221/13/07/P07006",
    journal = "J. Instrum{.}",
    volume = "13",
    number = "07",
    pages = "P07006",
    year = "2018"
}

@article{bib:uB_signal2,
    author = "Adams, C. and others",
    collaboration = "MicroBooNE Collaboration",
    title = "{Ionization electron signal processing in single phase LArTPCs. Part II. Data/simulation comparison and performance in MicroBooNE}",
    reportNumber = "FERMILAB-PUB-18-096-ND",
    doi = "10.1088/1748-0221/13/07/P07007",
    journal = "J. Instrum{.}",
    volume = "13",
    number = "07",
    pages = "P07007",
    year = "2018"
}

@article{mb:2007,
  title = "{Search for Electron Neutrino Appearance at the $\ensuremath{\Delta}{m}^{2}\ensuremath{\sim}1$ eV$^{2}$ Scale}",
  author = "Aguilar-Arevalo, A. A. and others",
  collaboration = {MiniBooNE Collaboration},
  journal = {Phys. Rev. Lett.},
  volume = {98},
  issue = {23},
  pages = {231801},
  numpages = {7},
  year = {2007},
  month = {6},
  publisher = {American Physical Society},
  doi = {10.1103/PhysRevLett.98.231801},
  url = {https://link.aps.org/doi/10.1103/PhysRevLett.98.231801}
}

@article{mb:2021,
    author = "Aguilar-Arevalo, A. A. and others",
    collaboration = "MiniBooNE Collaboration",
    title = "{Updated MiniBooNE neutrino oscillation results with increased data and new background studies}",
    reportNumber = "LA-UR-20-29235, LA-UR-20-24753, FERMILAB-PUB-20-288-AD-ND",
    doi = "10.1103/PhysRevD.103.052002",
    journal = "Phys. Rev. D",
    volume = "103",
    number = "5",
    pages = "052002",
    year = "2021"
}

@article{mb:2013,
    author = "Aguilar-Arevalo, A. A. and others",
    collaboration = "MiniBooNE Collaboration",
    title = "{Improved Search for $\bar \nu_\mu \rightarrow \bar \nu_e$ Oscillations in the MiniBooNE Experiment}",
    reportNumber = "FERMILAB-PUB-13-066-AD-E-PPD, LA-UR-13-21523",
    doi = "10.1103/PhysRevLett.110.161801",
    journal = "Phys. Rev. Lett.",
    volume = "110",
    pages = "161801",
    year = "2013"
}

@article{mb:2008,
    author = "Aguilar-Arevalo, A. A. and others",
    collaboration = "MiniBooNE Collaboration",
    title = "{Unexplained Excess of Electron-Like Events From a 1-GeV Neutrino Beam}",
    reportNumber = "FERMILAB-PUB-08-570, LA-UR-08-07142",
    doi = "10.1103/PhysRevLett.102.101802",
    journal = "Phys. Rev. Lett.",
    volume = "102",
    pages = "101802",
    year = "2009"
}

@article{mb:2018,
  title = "{Significant Excess of Electronlike Events in the MiniBooNE Short-Baseline Neutrino Experiment}",
  author = "Aguilar-Arevalo, A. A. and others",
  collaboration = {MiniBooNE Collaboration},
  journal = {Phys. Rev. Lett.},
  volume = {121},
  issue = {22},
  pages = {221801},
  numpages = {7},
  year = {2018},
  month = {11},
  publisher = {American Physical Society},
  doi = {10.1103/PhysRevLett.121.221801},
  url = {https://link.aps.org/doi/10.1103/PhysRevLett.121.221801}
}

@article{uB_gLEE,
    author = "Abratenko, P. and others",
    collaboration = "MicroBooNE Collaboration",
    title = "{Search for Neutrino-Induced Neutral-Current {$\mathrm{\ensuremath{\Delta}}$} Radiative Decay in MicroBooNE and a First Test of the MiniBooNE Low Energy Excess under a Single-Photon Hypothesis}",
    journal = {Phys. Rev. Lett.},
    volume = {128},
    issue = {11},
    pages = {111801},
    numpages = {8},
    year = {2022},
    month = {3},
    publisher = {American Physical Society},
    doi = {10.1103/PhysRevLett.128.111801},
    url = {https://link.aps.org/doi/10.1103/PhysRevLett.128.111801}
}

@article{uB_PeLEE,
  title = "{Search for an anomalous excess of charged-current ${\ensuremath{\nu}}_{e}$ interactions without pions in the final state with the MicroBooNE experiment}",
  author = {Abratenko, P. and others},
  collaboration = {MicroBooNE Collaboration},
  journal = {Phys. Rev. D},
  volume = {105},
  issue = {11},
  pages = {112004},
  numpages = {18},
  year = {2022},
  month = {6},
  publisher = {American Physical Society},
  doi = {10.1103/PhysRevD.105.112004},
  url = {https://link.aps.org/doi/10.1103/PhysRevD.105.112004}
}

@article{uB_WCeLEE,
  title = "{Search for an anomalous excess of inclusive charged-current ${\ensuremath{\nu}}_{e}$ interactions in the MicroBooNE experiment using Wire-Cell reconstruction}",
  author = {Abratenko, P. and others},
  collaboration = {MicroBooNE Collaboration},
  journal = {Phys. Rev. D},
  volume = {105},
  issue = {11},
  pages = {112005},
  numpages = {39},
  year = {2022},
  month = {6},
  publisher = {American Physical Society},
  doi = {10.1103/PhysRevD.105.112005},
  url = {https://link.aps.org/doi/10.1103/PhysRevD.105.112005}
}

@article{uB_DL,
  title = "{Search for an anomalous excess of charged-current quasielastic ${\ensuremath{\nu}}_{e}$ interactions with the MicroBooNE experiment using Deep-Learning-based reconstruction}",
  author = {Abratenko, P. and others},
  collaboration = {MicroBooNE Collaboration},
  journal = {Phys. Rev. D},
  volume = {105},
  issue = {11},
  pages = {112003},
  numpages = {39},
  year = {2022},
  month = {6},
  publisher = {American Physical Society},
  doi = {10.1103/PhysRevD.105.112003},
  url = {https://link.aps.org/doi/10.1103/PhysRevD.105.112003}
}

@article{uB_eLEE_PRL,
  title = "{Search for an Excess of Electron Neutrino Interactions in MicroBooNE Using Multiple Final-State Topologies}",
  author = {Abratenko, P. and others},
  collaboration = {MicroBooNE Collaboration},
  journal = {Phys. Rev. Lett.},
  volume = {128},
  issue = {24},
  pages = {241801},
  numpages = {9},
  year = {2022},
  month = {6},
  publisher = {American Physical Society},
  doi = {10.1103/PhysRevLett.128.241801},
  url = {https://link.aps.org/doi/10.1103/PhysRevLett.128.241801}
}

@article{Tena-Vidal:2021rpu,
    author = {Tena-Vidal, J\'ulia and others},
    collaboration = {GENIE},
    title = "{Neutrino-Nucleon Cross-Section Model Tuning in GENIE v3}",
    journal = {arXiv e-prints},
    eprint = "2104.09179",
    archivePrefix = "arXiv",
    primaryClass = "hep-ph",
    reportNumber = "FERMILAB-PUB-20-531-SCD-T",
    month = "4",
    year = "2021"
}

@article{generic_nu,
  title = "{Cosmic Ray Background Rejection with Wire-Cell LArTPC Event Reconstruction in the MicroBooNE Detector}",
  author = {Abratenko, P. and others},
  collaboration = {MicroBooNE Collaboration},
  journal = {Phys. Rev. Appl.},
  volume = {15},
  issue = {6},
  pages = {064071},
  numpages = {30},
  year = {2021},
  month = {6},
  publisher = {American Physical Society},
  doi = {10.1103/PhysRevApplied.15.064071},
  url = {https://link.aps.org/doi/10.1103/PhysRevApplied.15.064071}
}

@article{wire-cell-pr,
    author = "Abratenko, P. and others",
    collaboration = "MicroBooNE Collaboration",
    title = "{Wire-cell 3D pattern recognition techniques for neutrino event reconstruction in large LArTPCs: algorithm description and quantitative evaluation with MicroBooNE simulation}",
    reportNumber = "FERMILAB-PUB-21-509-ND",
    doi = "10.1088/1748-0221/17/01/P01037",
    journal = "J. Instrum{.}",
    volume = "17",
    number = "01",
    pages = "P01037",
    year = "2022"
}

@article{xgboost,
       author = {{Chen}, Tianqi and {Guestrin}, Carlos},
        title = "{XGBoost: A Scalable Tree Boosting System}",
      journal = "{arXiv e-prints}",
         year = {2016},
         month = {mar},
    archivePrefix = {arXiv},
       eprint = {1603.02754},
    primaryClass = {cs.LG},
    adsurl = {https://ui.adsabs.harvard.edu/abs/2016arXiv160302754C}
}

@article{CNP,
title = "{Combined Neyman–Pearson chi-square: An improved approximation to the Poisson-likelihood chi-square}",
journal = {Nucl. Instrum. Meth. A},
volume = {961},
pages = {163677},
year = {2020},
issn = {0168-9002},
doi = {https://doi.org/10.1016/j.nima.2020.163677},
url = {https://www.sciencedirect.com/science/article/pii/S0168900220302436},
author = {Xiangpan Ji and Wenqiang Gu and Xin Qian and Hanyu Wei and Chao Zhang}
}

@unpublished{bnb_flux,
note = { ``Booster Neutrino Flux Prediction at MicroBooNE'', MicroBooNE public-note 1031, \url{http://microboone.fnal.gov/wp-content/uploads/MICROBOONE-NOTE-1031-PUB.pdf}}
}

@article{Qian:2018qbv,
    author = "Qian, Xin and Zhang, Chao and Viren, Brett and Diwan, Milind",
    title = "{Three-dimensional Imaging for Large LArTPCs}",
    doi = "10.1088/1748-0221/13/05/P05032",
    journal = "J. Instrum{.}",
    volume = "13",
    number = "05",
    pages = "P05032",
    year = "2018"
}

@article{Abratenko:2020hpp,
    author = "Abratenko, P. and others",
    collaboration = "MicroBooNE Collaboration",
    title = "{Neutrino event selection in the MicroBooNE liquid argon time projection chamber using Wire-Cell 3D imaging, clustering, and charge-light matching}",
    reportNumber = "FERMILAB-PUB-20-578-ND",
    doi = "10.1088/1748-0221/16/06/P06043",
    journal = "J. Instrum{.}",
    volume = "16",
    number = "06",
    pages = "P06043",
    year = "2021"
}

@article{bayes,
	title = "{Bayesian statistics and modelling}",
	volume = {1},
	copyright = {2021 Springer Nature Limited},
	issn = {2662-8449},
	url = {https://www.nature.com/articles/s43586-020-00001-2},
	doi = {10.1038/s43586-020-00001-2},
	number = {1},
	urldate = {2024-04-07},
	journal = {Nat. Rev. Methods Primers},
	author = {van de Schoot, Rens and Depaoli, Sarah and King, Ruth and Kramer, Bianca and Märtens, Kaspar and Tadesse, Mahlet G. and Vannucci, Marina and Gelman, Andrew and Veen, Duco and Willemsen, Joukje and Yau, Christopher},
	month = {1},
	year = {2021},
	pages = {1--26},
}

@article{Miniboonedetector,
    author = "Aguilar-Arevalo, A. A. and others",
    collaboration = "MiniBooNE Collaboration",
    title = "{The MiniBooNE Detector}",
    reportNumber = "FERMILAB-PUB-08-210",
    doi = "10.1016/j.nima.2008.10.028",
    journal = "Nucl. Instrum. Meth. A",
    volume = "599",
    pages = "28--46",
    year = "2009"
}

@article{pelee15,
  title = "{Search for an Anomalous Production of Charged-Current ${\ensuremath{\nu}}_{e}$ Interactions without Visible Pions across Multiple Kinematic Observables in MicroBooNE}",
  collaboration = {MicroBooNE collaboration},
      author={P. Abratenko and others},
  journal = {Phys. Rev. Lett.},
  volume = {135},
  issue = {8},
  pages = {081802},
  numpages = {8},
  year = {2025},
  month = {Aug},
  publisher = {American Physical Society},
  doi = {10.1103/x259-z6mf},
  url = {https://link.aps.org/doi/10.1103/x259-z6mf}
}

@article{bib:ben_numuCC,
  title = "{First Simultaneous Measurement of Differential Muon-Neutrino Charged-Current Cross Sections on Argon for Final States with and without Protons Using MicroBooNE Data}",
  author = {Abratenko, P. and others},
  collaboration = {MicroBooNE Collaboration},
  journal = {Phys. Rev. Lett.},
  volume = {133},
  issue = {4},
  pages = {041801},
  numpages = {8},
  year = {2024},
  month = {Jul},
  publisher = {American Physical Society},
  doi = {10.1103/PhysRevLett.133.041801},
  url = {https://link.aps.org/doi/10.1103/PhysRevLett.133.041801}
}

@article{blipreco,
  title = "{Demonstration of new MeV-scale capabilities in large neutrino LArTPCs using ambient radiogenic and cosmogenic activity in MicroBooNE}",
  author = {Abratenko, P. and others},
  collaboration = {MicroBooNE Collaboration},
  journal = {Phys. Rev. D},
  volume = {111},
  issue = {3},
  pages = {032005},
  numpages = {17},
  year = {2025},
  month = {Feb},
  publisher = {American Physical Society},
  doi = {10.1103/PhysRevD.111.032005},
  url = {https://link.aps.org/doi/10.1103/PhysRevD.111.032005}
}

@article{NCCohPaper,
      title="{First Search for Neutral Current Coherent Single-Photon Production in MicroBooNE}", 
      author = {Abratenko, P. and others},
      collaboration = {MicroBooNE Collaboration},
      year={2025},
      eprint={2502.06091},
      archivePrefix={arXiv},
      primaryClass={hep-ex},
      url={https://arxiv.org/abs/2502.06091}, 
      journal = {arXiv e-prints}
}

@article{2DNCDelPaper,
  title = "{Enhanced search for neutral current $\mathrm{\ensuremath{\Delta}}$ radiative single-photon production in MicroBooNE}",
  author = {Abratenko, P. and others},
  collaboration = {The MicroBooNE Collaboration},
  journal = {Phys. Rev. D},
  volume = {112},
  issue = {9},
  pages = {L091101},
  numpages = {9},
  year = {2025},
  month = {Nov},
  publisher = {American Physical Society},
  doi = {10.1103/49ds-5hfh},
  url = {https://link.aps.org/doi/10.1103/49ds-5hfh}
}

@phdthesis{mythesis,
    author = "Yandel, Erin Haley",
    title = "{A Search for Anomalous Single Photon Production in Neutrino Interactions with the MicroBooNE Detector}",
    school = "UC Santa Barbara",
    year = "2024"
}

@article{muonrealrad,
  title = "{Theory of QED radiative corrections to neutrino scattering at accelerator energies}",
  author = "Tomalak, Oleksandr and Chen, Qing and Hill, Richard J. and McFarland, Kevin S. and Wret, Clarence",
  journal = "{Phys. Rev. D}",
  volume = {106},
  issue = {9},
  pages = {093006},
  numpages = {35},
  year = {2022},
  month = "{Nov}",
  publisher = "{American Physical Society}",
  doi = {10.1103/PhysRevD.106.093006},
  url = {https://link.aps.org/doi/10.1103/PhysRevD.106.093006}
}

@misc{inclusive_hepdata,
    author = {Abratenko, P. and others},
    collaboration = "{MicroBooNE Collaboration}",
    title = "{Inclusive Search for Anomalous Single-Photon Production in MicroBooNE}",
    howpublished = "{HEPData (collection)}",
    year = 2025,
    note = "\url{https://doi.org/10.17182/hepdata.158440}"
}
